\documentclass[letterpaper, 10 pt, conference]{ieeeconf}    
\IEEEoverridecommandlockouts 
\usepackage[utf8]{inputenc}
\usepackage{amsmath}
\usepackage{amssymb}
\usepackage[font=small]{subfig}
\usepackage[font=small]{caption}
\usepackage{amsfonts}
\usepackage{mathtools}
\usepackage{url}
\usepackage{bm}
\usepackage{enumerate}
\usepackage{xcolor}
\usepackage{graphicx}
\usepackage{color}
\usepackage{tikz}
\usepackage{booktabs}
\usetikzlibrary{positioning, arrows, calc}

\let\labelindent\relax
\usepackage{enumitem}
\setlist[enumerate]{leftmargin=*}
\setlist[itemize]{leftmargin=*}

\usepackage[
    style=numeric,
    sorting=none,        
    sortcites=true,      
    backend=biber, 
    maxbibnames=99,      
    natbib=true,
    doi=false,   
    url=false,   
    isbn=false,  
    eprint=false, 
    giveninits=true
]{biblatex}

\newtheorem{prob}{Problem}
\newtheorem{remark}{Remark}

\title{\LARGE \bf Trajectory Control of the Suspended Load Pose using\\ Non-stopping Flying Carriers}
\author{Sofia Girardello$^1$, Giulia Michieletto$^{1,2}$, Angelo Cenedese$^{1,3}$, Antonio Franchi$^{4,5}$, Chiara Gabellieri$^4$ 
\thanks{$^1$Dept. of Information Engineering, University of Padova, Italy.
}
\thanks{$^2$Dept. of Management and  Engineering, University of Padova, Italy.
}
\thanks{$^3$Dept. of Industrial Engineering, University of Padova, Italy.
}
\thanks{$^4$Robotics and Mechatronics Dept., Electrical Engineering,  Mathematics, and Computer Science Faculty, University of Twente, The Netherlands. 
}
\thanks{$^5$Dept. of Computer, Control and Management Engineering, Sapienza University of Rome, Italy.
}
\thanks{Authors' contacts: 
C.G.:~{\tt\footnotesize c.gabellieri@utwente.nl}, 
A.F.:~{\tt\footnotesize schol@r-franchi.eu}, 
A.C.:~{\tt\footnotesize angelo.cenedese@unipd.it}, 
G.M.:~{\tt\footnotesize giulia.michieletto@unipd.it}. 
}
\thanks{The work was partially funded by the European Commission Horizon
Europe Framework under project Autoassess (101120732)}
}

\begin{document} 

\maketitle
\begin{abstract}
This work presents the first \textit{closed-loop} control framework for cooperative payload transportation with \textit{non-stopping flying carriers}. The proposed method includes a feedback wrench-controller that actively regulates the load’s pose by computing the wrench required for tracking its desired pose trajectory. Building upon grasp-matrix formulation and internal force redundancy, an optimization layer dynamically shapes internal-force parameters to guarantee persistent carrier motion, while not altering the desired load wrench. The desired non-stopping carrier's trajectories are computed using the system's kinematics and desired cable forces. Numerical simulations demonstrate that the method successfully prevents the carriers from stopping, while achieving a successful tracking of the desired load trajectory.
\end{abstract}

\section{Introduction}

The transportation of cable-suspended payloads using Uncrewed Aerial Vehicles (UAVs) has emerged as a key research direction in aerial robotics, driven by applications in long-range logistics, humanitarian relief, infrastructure inspection, and construction~\cite{estevez2024review}. In particular, cable-suspended payload manipulation offers several advantages over rigid grasping, including mechanical simplicity, reduced weight, and limited impact on the carrier attitude dynamics when cables are attached near the center of mass.

\subsubsection*{Related Literature} Most existing works on cable-suspended payload manipulation entail the use of multirotors as flying carriers, exploiting their hovering capabilities and precise low-speed maneuverability. These platforms can regulate cable tensions through thrust control, allowing the payload to be stabilized at a static equilibrium.
Several studies have demonstrated the feasibility of \textit{single-multirotor carrier systems}, where the main challenge is the suppression of pendulum-like oscillations through trajectory shaping or feedback control~\cite{8825990, Pereira2016, Sreenath2013}. Nonetheless, such systems are limited in payload capacity, robustness, and control authority over the pose of the load.
To overcome these limitations, \textit{cooperative multirotor carrier systems} have been proposed, where multiple platforms collaborate to enable distributed lifting and six-degree-of-freedom control of a cable-suspended load~\cite{michael2011cooperative, Masone2016, Chen2019, Sanalitro2020,li2021cooperative, gabellieri2023equilibria}. Although effective, these solutions critically rely on the assumption that the carriers can hover (stop) in mid air. 

Despite their versatility, multirotors are inherently constrained by limited endurance and short operational ranges, which restrict their scalability to long-distance transport. In contrast, fixed-wing UAVs offer superior aerodynamic efficiency, range, and endurance~\cite{leutenegger2016flying}, which are key features for object transportation and delivery~\cite{saengphet2016conceptual, zhang2024design}.
However, these vehicles introduce a fundamental challenge: unlike multirotors, they cannot hover and must maintain a strictly positive forward speed to remain airborne. 

Accounting for this requirement, in \cite{williams2009dynamics}, transport of a \textit{point-mass} load tethered to two \textit{cooperative non-stoppping fixed-wing carriers} has been theorized, and the trajectories for the vehicles have been planned, while \cite{quenneville2023experimental} experimentally showed load lifting for the same system.
More recently, theoretical advances began to address cooperative manipulation of \textit{rigid bodies}. 
In particular,~\cite{gabellieri2024existence} and~\cite{gabellieri2025coordinated} establish the feasibility of maintaining a static equilibrium of a cable-suspended load under continuous motion of three or more non-stopping carriers, and propose a coordinated trajectory generation method that guarantees non-zero velocity for the UAVs. These contributions provide the foundations for trajectory planning but remain limited to static loads and open-loop execution.
\begin{figure}[t]
    \centering
\includegraphics[width=\linewidth, trim = {4.7cm, 4cm, 7.5cm, 2.2cm}, clip]{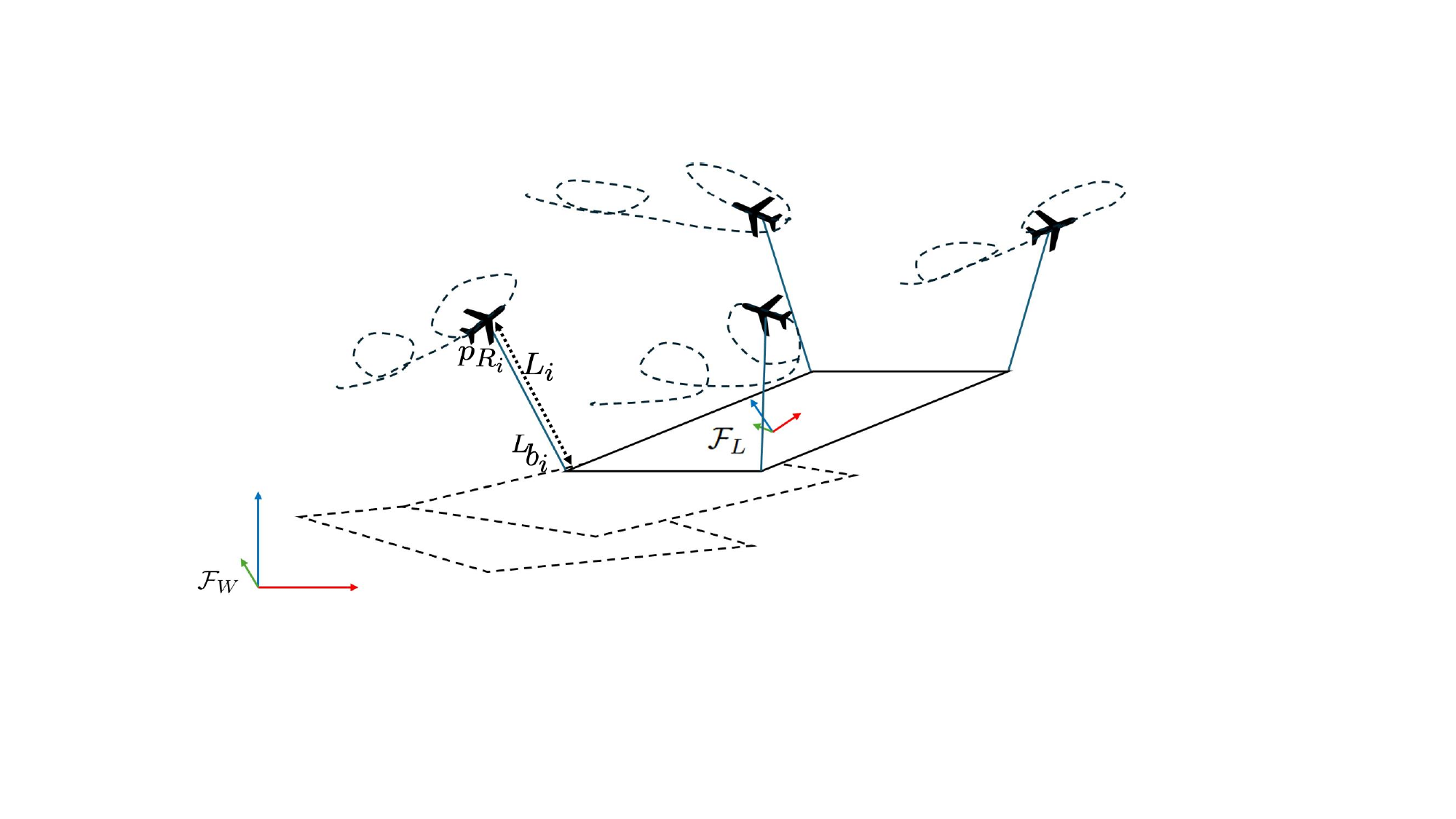}
    \caption{Trajectory tracking of the suspended load full (6D) pose using non-stopping flying carriers.}
    \label{fig:one}
\end{figure}
\medskip

\subsubsection*{Contributions of This Work }

To the best of the authors' knowledge, this work proposes the first control architecture for non-stopping flying carriers that enables tracking of dynamic load trajectories while integrating non-stopping-carrier constraints.
More in detail, the main contributions of the paper involve
\begin{itemize}
    \item the design of a feedback wrench controller to regulate the payload’s pose;
    \item the formulation of an online optimization of the system's internal forces to enforce non-zero carrier velocities and maintain the load trajectory tracking.
\end{itemize}
The proposed framework is evaluated by numerical simulations, demonstrating its ability to control the load trajectory, guaranteeing non-vanishing velocities of the carriers for trajectories of the load that embed dynamic and rest phases.

\section{System Model}\label{sec:model}

We consider a cooperative aerial transportation system composed of $n \geq 3$ non-stopping aerial vehicles, each connected via a cable to a common rigid payload and required to maintain a strictly non-zero velocity norm to satisfy aerodynamic constraints. Refer to Figure~\ref{fig:one} for a schematic representation of the considered scenario. 

We assume that 
$(i)$ the cables are massless and inextensible  and tensioned during the task execution; 
$(ii)$ the payload acts as a rigid body (dynamic modeling) while the carriers act as (massless) frames (kinematic modeling).

To describe the system, we consider the inertial world frame $\mathcal{F}_W=\{O_W, (x_W, y_W, z_W)\}$, with gravity acting along the opposed direction of $z_W$-axis, and the body-fixed frame $\mathcal{F}_L$  attached to the center of mass (CoM) of the load, which has mass $m_L \in \mathbb{R}_{>0}$ and inertia matrix $J_L \in \mathbb{R}^{3 \times 3}$. 
{The $i$-th flying carrier, $i = 1, \ldots, n$,} is connected to a fixed attachment point $B_i$ on the payload via a cable of constant length $L_i \in \mathbb{R}_{>0}$, with unit direction vector $q_i{(t)} \in \mathbb{S}^2$ pointing from the payload to the carrier.  

The position $p_{R_i}{(t)} \in \mathbb{R}^3$ of the $i$-th carrier expressed in $\mathcal{F}_W$ is thus constrained by the payload configuration:
\begin{equation}
p_{R_i}{(t)} = p_L{(t)} + R_L{(t)}\,{}^L\!b_i + L_i\,q_i{(t)}, 
\label{eq:carrierpos}
\end{equation}
where $p_L{(t)} \in \mathbb{R}^3$ and $R_L{(t)} \in \mathrm{SO}(3)$ denote the payload position and orientation in $\mathcal{F}_W$, and ${}^L\!b_i \in \mathbb{R}^3$ represent the cable attachment points expressed in $\mathcal{F}_L$ (see Figure~\ref{fig:one}). 
Differentiating~\eqref{eq:carrierpos} yields {the carrier's kinematics equation:}
\begin{equation}
\dot{p}_{R_i}{(t)} = \dot{p}_L{(t)} + \dot{R}_L{(t)}\,{}^L\!b_i + L_i\,\dot{q}_i{(t)},\label{eq:carrier_vel_0}
\end{equation}
with $\dot R_L{(t)} = R_L{(t)} S(^L\omega_L{(t)})$ where  $^L\omega_L{(t)} \in \mathbb{R}^3$ is the payload angular velocity w.r.t. $\mathcal{F}_W$, expressed in $\mathcal{F}_L$,\footnote{Quantities are considered expressed in the world frame unless otherwise indicated by the left superscript.} and $S(\cdot)$ is the skew-symmetric operator.  Eq.~\eqref{eq:carrier_vel_0} shows that each UAV velocity is inherently constrained by the payload motion and the cable geometry.
Moreover, the non-stopping carrier dynamics impose  the following \textit{minimum-norm velocity constraints} {where $\varepsilon \in \mathbb{R}_{>0}$ is a lower-bound threshold}:
\begin{equation}
\label{eq:min_velocity}
\|\dot p_{Ri}(t)\| \ge \varepsilon > 0. 
\end{equation}

The tension $T_i(t) \in \mathbb{R}_{\geq 0}$ in the $i$-th cable generates a force $f_i(t) \in \mathbb{R}^3$ on the payload along the direction $q_i$, namely,
\begin{equation}
f_i(t) = T_i(t) q_i(t) \label{eq:cable_tension_direction}
\end{equation}
The payload dynamics follows the Newton–Euler equations:
\begin{align}
m_L \ddot p_L(t) &= -m_L g e_3 + \sum_{i=1}^n f_i(t), \label{eq:load_trans}\\
J_L ^L\dot \omega_L(t) &= -^L\omega_L(t) \times J_L {^L\omega_L(t)} + \sum_{i=1}^n S({}^L\!b_i)\,R_L(t)^\top f_i(t). \label{eq:load_rot}
\end{align}

We introduce $f(t) = \left[f_1(t)^\top \;\dots \; f_n(t)^\top\right]^\top \in \mathbb{R}^{3n}$ collecting the cable forces, and $w(t) = \left[f_L(t)^\top\; \tau_L(t)^\top\right]^\top \in \mathbb{R}^6$ representing the wrench at the load CoM, where $f_L(t) \in \mathbb{R}^3$ and $\tau_L(t) \in \mathbb{R}^3$ denote the force and the moment components, respectively. The relation between these vectors is given by
\begin{equation}
w(t) = G(R_L(t)) f(t), \label{eq:wrench}
\end{equation}
where $G(R_L(t)) \in \mathbb{R}^{6 \times 3n}$ is the grasp matrix defined as
\begin{equation}
G(R_L(t)) =
\begin{bmatrix}
I_3 & \cdots & I_3 \\
S(R_L(t){}^L\!b_1) & \cdots & S(R_L(t){}^L\!b_n))
\end{bmatrix}.
\end{equation}
Trivially, the first row-block maps cable forces $f$ to translational force $f_L(t)$, while the second row-block maps them to the moment $\tau_L(t)$ about the payload CoM. 
Eq.~\eqref{eq:wrench} can be used to figure out the cable forces that exert the resultant wrench $w$ on the load. Indeed, we have that
\begin{equation}
f(t) =  
\underbrace{G(R_L(t))^\dagger w}_{=:f_g(t)} + 
\underbrace{N(R_L(t))\lambda(t)}_{ =:f_\lambda(t)},\label{eq:cable_forces}
\end{equation}
where $(\bullet)^\dagger$ indicates the right pseudo inverse and $N(R_L(t))$ is {a matrix whose columns identify a} basis of the nullspace of $G(R_L(t))$. 
Under the mild  assumption that the attachment points on the load guarantee that $\operatorname{rank}(G(R_L(t)))=6$ we obtain that the dimension of $\operatorname{null}(G(R_L(t)))$ is $3n - 6$ and is equal to the number of columns of $N(R_L(t))$. This corresponds to the number of independent internal force components--i.e., the degree of redundancy--available in the system. In particular, system redundancy allows for \textit{internal forces}  $f_\lambda(t) \in \mathbb{R}^{3n}$ belonging to $\operatorname{null}(G(R_L(t)))$, which enable the redistribution of cable tensions \textit{without affecting the load wrench, and hence the load motion}. As done in \cite{gabellieri2024existence}, in this work we operate a choice of $N(R_L(t))$ to parametrize these internal forces by a set $\lambda(t)$ of $n$ functions.

\section{Problem Statement}

For the system described in Section \ref{sec:model},
a desired load trajectory is specified by the desired position,  orientation, and velocities over time, indicated as $p_{Ld}(t) \in \mathbb{R}^3$, $ R_{Ld}(t) \in \mathrm{SO(3)}$, and $\dot{p}_{Ld}(t), ^L\omega_{Ld}(t) \in \mathbb{R}^3$. The pose and velocity errors of the load w.r.t. the desired trajectory are defined as follows:
\begin{align}
&e_p(t) = p_L(t) - p_{Ld}(t)\label{eq:pos_errors}\\ 
&e_R(t)=\frac{1}{2}
(R_{Ld}(t)^\top R_L(t)-R_{L}(t)^\top R_{Ld}(t))^\vee
\\
&e_v(t)=\dot{p}_L(t)-\dot{p}_{Ld}(t) \\ &e_\omega(t)=^L\omega_{Ld}(t)-^L\omega_L(t),\label{eq:vel_errors}
\end{align}
where $(\cdot)^\vee$ is the inverse of the skew-symmetric operator.

\begin{prob} Find cable force trajectories and corresponding carrier position trajectories 
$
\{f_i(t), p_{Ri}(t)\}_{i=1}^n
$
such that

\begin{enumerate}
    \item[$i$)] 
the load tracking errors~\eqref{eq:pos_errors}-\eqref{eq:vel_errors} are regulated and remain bounded, converging to zero for constant references;

    \item[$ii)$] the non-stopping carrier minimum-norm velocity constraints \eqref{eq:min_velocity} and load dynamics \eqref{eq:load_trans}-\eqref{eq:load_rot} are satisfied.
\end{enumerate}
\end{prob}

\begin{remark}
In~\cite{gabellieri2024existence, gabellieri2025coordinated} it is established that a load can be statically supported by a minimum of three non-stopping flying carriers and an open-loop strategy is proposed to control a cable-suspended \textit{static} load through non-stopping carriers that exploit the nullspace of the grasp matrix to generate time-varying internal forces in the cables.
More in detail, in \cite{gabellieri2025coordinated}, a suitable nullspace basis of the grasp matrix is exploited and sinusoidal functions having constant amplitude $A \in \mathbb{R}$, frequency $\xi \in \mathbb{R}$, and phase shifts $\phi_i \in [-\pi, \pi)$, $i=1,\ldots,n$ are assigned to $\lambda(t)$ to generate open-loop carrier trajectories that ensure persistent carriers' motion for a given static load pose. Notably, when dealing with static load, the grasp matrix is constant over time since the load orientation is fixed.\end{remark}

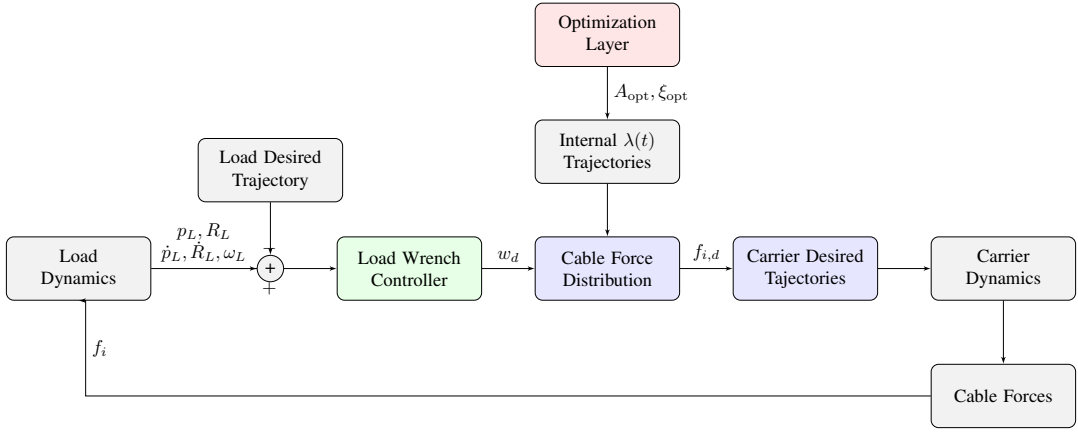
\begin{figure*}[t!]
\centering
\resizebox{0.8\textwidth}{!}{%
\begin{tikzpicture}[auto, node distance=2.2cm,>=latex', font=\normalsize]

\tikzstyle{block} = [rectangle, draw, fill=gray!10, 
    text width=2.5cm, text centered, rounded corners, minimum height=1.2cm]
\tikzstyle{optblock} = [rectangle, draw, fill=red!10, 
    text width=2.5cm, text centered, rounded corners, minimum height=1.2cm]
\tikzstyle{greenblock} = [rectangle, draw, fill=green!10, 
    text width=2.5cm, text centered, rounded corners, minimum height=1.2cm]
\tikzstyle{blueblock} = [rectangle, draw, fill=blue!10, 
    text width=2.5cm, text centered, rounded corners, minimum height=1.2cm]
\tikzstyle{sum} = [circle, draw, inner sep=0mm, minimum size=5mm, fill=gray!10]
%
\node [block] (load) {Load \\Dynamics};
\node [sum, right=2cm of load] (sum1) {+};
\node [greenblock, right=1cm of sum1] (wrench) {Load Wrench \\Controller};
\node [block, above=1cm of sum1] (des_traj) {Load Desired Trajectory};
\node [blueblock, right=1cm of wrench] (allocation) {Cable Force \\Distribution};
\node [blueblock, right=1cm of allocation] (endpoints) {Carrier Desired Tajectories};
\node [block, above=1cm of allocation] (traj) {Internal $\lambda(t)$ Trajectories};
\node [optblock, above=1cm of traj] (opt) {Optimization \\ Layer};
\node [block, right=1cm of endpoints] (carrier) {Carrier Controller\\ + Dynamics};
\node [block, below=1.2cm of carrier] (forcecomp) {Cable Forces};
%
\draw[->] (load.east) -- ++(2.0,0) node[midway, above, align=center] {%
    $p_L, R_L$ \\ $\dot p_L, \dot R_L$} -- (sum1.west);
   \draw[->] (load.north) |- (opt.west) node[pos=0.8, above] {%
    $p_L, R_L$  $\dot p_L, \dot R_L$} ;
    \draw[->] (wrench.north) |- (opt.west) node[pos=0.8, above, align=center] {$w_d$} ;
\draw[->] (des_traj.south) -- (sum1.north);
\draw[->] (sum1.east) -- (wrench.west);
\draw[->] (wrench.east) -- node[midway, above, align=center] {$w_d$} (allocation.west);
\draw[->] (allocation.east) -- node[midway, above, align=center] {$f_{i,d}$} (endpoints.west);
\draw[->] (endpoints.east) -- (carrier.west);
\draw[->] (opt.south) -- (traj.north) 
    node[pos=0.5, right, align=center] {%
     $A^*, \xi^*$};

\draw[->] (traj.south) -- (allocation.north) node[pos=0.5, right, align=center] {%
     $\lambda*$};
\draw[->] (carrier.south) -- node[midway, right] {} (forcecomp.north);
\draw[->] (forcecomp.west) -| (load.south) 
    node[pos=0.25, above] {$f_i$};
%
\node at (sum1.north east) [xshift=-2mm,yshift=2mm] {$-$};

\end{tikzpicture}%
}
\caption{Overall control scheme including the outer-loop load-wrench controller (green block), the inner-loop carrier-trajectory generator (blue blocks), and the optimization layer (red block).}
\label{fig:full_feedback_loop_opt}
\end{figure*}

\section{Proposed Solution}
Moving beyond the open-loop strategy presented in~\cite{gabellieri2025coordinated}, we propose an optimization-based \textit{feedback} controller that regulates the load’s pose along a desired trajectory while ensuring persistent motion of the non-stopping flying carriers. Similar to \cite{gabellieri2025coordinated}, sinusoidal functions are assigned to the internal force parameters $\lambda(t)$. However, the proposed controller changes online the amplitude and frequency of the sinusoidal functions to satisfy the minimum-norm velocity constraints of the carriers.

\begin{remark}
     For a moving load, it may seem reasonable to just let the non-stopping vehicles `follow' the motion of the load with a displacement provided by the kinematics in~\eqref{eq:carrierpos}. However, this work considers the more challenging and interesting scenario in which the desired load trajectory may exhibit a velocity with an arbitrarily small norm and ultimately even stop during the mission (e.g., for obstacle avoidance, or requirements of the task at hand). Hence, this work considers elliptical trajectories of the carriers, enabled by imposing sinusoidal internal forces, as in~\cite{gabellieri2025coordinated}. Those trajectories ensure that the carriers can keep moving even when the load must stop or slow down. 
\end{remark}


Figure~\ref{fig:full_feedback_loop_opt} shows the high-level architecture of the proposed method, combining:
\begin{itemize}
    \item an \textbf{outer-loop load-wrench controller} 
    computing the desired wrench $w_d(t) \in \mathbb{R}^6$ on the load;
    \item an \textbf{optimal internal force  generator}
    designing the internal forces $f_\lambda(t)$ to enforce a strictly positive carrier velocity norm.   
    \item a \textbf{carrier-trajectory generator} 
    converting $w_d(t)$ and the internal forces from the previous two into cable tensions and carrier trajectories, thus exploiting the overall redundancy of the system. 

\end{itemize}

Therefore, the proposed architecture separates load tracking from carrier trajectory feasibility. The load wrench controller computes the wrench required for load pose regulation. The cable-force distribution realizes this wrench while the online optimization layer updates only the internal force parameters so that carrier velocities satisfy the non-stopping requirement, while the commanded load wrench remains unchanged.

\subsection{Outer-Loop Load-Wrench Controller}
Given the desired load trajectory, and the errors in \eqref{eq:pos_errors}, \eqref{eq:vel_errors}, 
this module computes the desired load wrench $w_d(t)=\left[ f_{L,d}(t)^\top \; \tau_{L,d}(t)^\top \right]^\top$
using PID controllers, namely
\begin{align}
f_{L,d}(t) &=  m_L g e_3+ \label{eq:PID_force}\\
&\qquad -K_p e_p(t) - K_v e_v(t) - K_i \int_0^t e_p(\tau)d\tau ,\nonumber\\
^L\tau_{L,d}(t) &= ^L\omega_L(t) \times J_L {^L\omega_L(t)}+ \label{eq:PID_torque}  \\
& \qquad -K_R e_R(t) - K_\omega e_\omega(t) - K_{iR} \int_0^t e_R(\tau)d\tau , \nonumber
\end{align}
with tunable gains $K_p, K_v, K_i,  K_R, K_\omega, K_{iR}  \in \mathbb{R}^{3 \times 3}$.

\subsection{Optimization layer: Optimal Internal Force  Parameters}

By construction, the internal cable force components $f_\lambda(t)$ in~\eqref{eq:cable_forces} do not affect the total wrench at the load’s CoM.
In this work, we propose to modulate these forces online, solving an optimization problem in order to keep the carrier velocities above a desired threshold.

The first step to define the optimization problem is to express the UAV velocities as a function of the current state and the desired wrench given by the load-wrench controller.
Given $w_d(t)$ based on~\eqref{eq:PID_force}-\eqref{eq:PID_torque}, 
and
the corresponding cable forces $f(t)\in\mathbb{R}^{3n}$ generating such a wrench is:
\begin{equation}
f(t) = G(t)^\dagger w_d(t) + N(t)\,\lambda(t),\label{eq:desired_cable_forces}
\end{equation}
where $\lambda(t)$ represents a functional  variable to be optimized, 
and we use the short hand $G(t)=G(R_L(t))$ and ${N(t)=N(R_L(t))}$.

We derive the conditions required to comply with the non-stopping-vehicle constraint~\eqref{eq:min_velocity}. The carrier trajectories are dependent on the load trajectory via~\eqref{eq:carrierpos}-\eqref{eq:carrier_vel_0}. 
Introducing ${v_{Li}(t) = \dot p_L(t) + \dot R_L(t){}^B\!b_i} \in \mathbb{R}^3
$ and accounting for~\eqref{eq:cable_tension_direction}, the $i$-th carrier velocity can be rewritten as
\begin{equation}
{\dot p_{R_i}(t) = v_{Li}(t) + L_i \frac{1}{T_i(t)} \dot f_i^\perp(t)},
\end{equation}
where ${\dot f_i^\perp(t) = \Pi_i(t)\dot f_i(t)} \in \mathbb{R}^3$ is the projection of the cable force derivative orthogonal to $q_i(t)$, based on the projector matrix 
\begin{equation}
\Pi_i(t) = I_3 - q_i(t) q_i^\top(t) \in \mathbb{R}^{3 \times 3}.
\end{equation}
{Indeed, from \eqref{eq:cable_tension_direction} it follows that $\dot{f}_i(t)=T_i(t)\dot{q}_i(t)+\dot{T}_i(t)q_i(t)$ and, considering that $\dot{q}_i^\top(t) q_i(t)=0$, that $\dot{q}_i(t)=\frac{{\dot{f}_i}^\perp(t)}{T_i(t)}
$.}
{Based on~\eqref{eq:cable_forces}}, the derivative of each cable force can be decomposed as
\begin{equation}
\dot f_i(t) = e_i(t) + g_i(t),
\end{equation}
where the term $e_i(t) \in \mathbb{R}^3$ collects external contributions due to payload acceleration and angular velocity dynamics, and the term $g_i(t) \in \mathbb{R}^3$ collects internal contributions due to $\lambda(t)$ and $ \dot\lambda(t)$. More explicitly, we have that
\begin{align}
e(t) &= \dot G(t)^\dagger w(t) + G(t)^\dagger \dot w(t),\\
g(t) &= \dot N(t)\lambda(t) + N(t)\dot\lambda(t).
\end{align}
Hence, it follows that $\dot f_i^\perp(t) = \Pi_i(t)e_i(t) + \Pi_i(t)g_i(t)$, 
%
and the carrier velocity is expressed as
 \begin{equation}
 \dot p_{R_i}(t) = v_{Li}(t) + \frac{L_i}{T_i(t)}\big(\Pi_i(t)e_i(t) + \Pi_i(t)g_i(t)\big).\label{eq:carrier_vel}
 \end{equation}

Thus, the non-stopping requirement in \eqref{eq:min_velocity} with ${\varepsilon=0}$
is equivalent to excluding the origin from the set ${\mathcal{V}_i = \big\{ \dot p_{R_i}(t) \text{\; as in~\eqref{eq:carrier_vel}},\, t \ge 0 \big\}}$;  conversely, if the origin is reached, the non-stopping constraint is violated.

This provides an explicit mathematical constraint on the admissible internal force trajectories, leading to the optimization described in the following.

Choosing $\lambda(t)$ as a set of $n$ time-dependent  sinusoidal functions:
\begin{align}
\begin{aligned}
\lambda_i(\xi(t),A(t),t) &= A(t) \cos(\xi(t) t + \phi_i),\\
\dot\lambda_i(\xi(t),A(t),t) &= -A(t)\xi(t) \sin(\xi(t) t + \phi_i),
\end{aligned}
\label{eq:lambdas}
\end{align}
where the parameters $\phi_i \in [-\pi, \pi)$, $i=1,\ldots,n$ are constant and selected as in~\cite{gabellieri2025coordinated}, whereas the pair $x(t)=(\xi(t), A(t))$, contrary to the state of the art, is optimized online to enforce a desired lower-bound on the norm of the carrier velocities. Formally, 
given $w_d(t)$ based on~\eqref{eq:PID_force}-\eqref{eq:PID_torque}, 
for any $i$-th carrier, $i=1,...,n$ the constraint to exclude the origin from $\mathcal{V}_i$ is 
\begin{equation}
\Bigl\| v_{Li}(t) + \frac{L_i}{T_i(t)}\,\Pi_i(t)\big(\dot G_i^\dagger w_d + G_i^\dagger \dot w_d + \dot N_i \lambda + N_i \dot\lambda \big)\Bigr\|_2 \ge \varepsilon,\label{eq:norm_epsilon}
\end{equation}
with $\varepsilon>0$, and where we indicate with $\dot G_i^\dagger $ and $ N_i$ the $3\times 6$ and $3\times n$ row-blocks of $G^\dagger$ and $N$, respectively, that refer to the $i^{th}$ carrier force. 

The optimization problem solved online is 
\begin{equation}
\begin{aligned}
x^\star(t) = \arg\min_{x(t)\in\mathbb{R}^2} \;& J(x(t)) \\
\text{s.t. \, eq.}&  \eqref{eq:norm_epsilon} \quad \forall i=1,\dots,n,
\end{aligned}
\end{equation}
with the objective function 
\begin{equation}
\begin{aligned}
{J(x(t))} = & (\xi(t)-\xi(t^-))^2 + (A(t)-A(t^-))^2 \\
       & + w_{\text{pos}} \, \|\lambda(x(t))-\lambda(x(t^-))\|_2^2 \\
       & + w_{\text{vel}} \, \|\dot\lambda(x(t))-\dot\lambda(x(t^-))\|_2^2.
\end{aligned}
\end{equation}
where {$t^-$} indicates the value at the previous time instant and $w_{\text{pos}}, w_{\text{vel}}$ are positive weights. The cost function is chosen to discourage instantaneous large variations of the cable forces, and so potential large variations in the carriers' trajectories. 

Solving the optimization results in the optimal internal  force functional coefficients $\lambda_i(\xi^*(t),A^*(t),t)=:\lambda_i^*(t)$  for $i=1,\ldots,n$.

\subsection{Carrier-Trajectory Generator}
The cable force distribution computes the desired cable forces $f_{i,d}$ using \eqref{eq:desired_cable_forces} with $\lambda(t)=\lambda^*(t).$ Hence, the carrier trajectory generation module computes the carriers' trajectories as 
$${p_R}_{i,d}(t) = p_{Ld}(t) + R_{Ld}(t){^Bb_i}+L_i q_{i,d},$$ where $q_{i,d}$ is computed as $f_{i,d}/||f_{i,d}||_2$.  $\dot{p}_{R_{i,d}}(t)$ is computed similarly.
These references are tracked by the carriers' low-level position controller.

\subsection{Remark on the Implementation}

This framework ensures the persistent motion of all carriers while maintaining load-tracking. Note that the optimization does not change those components of the cable forces that generate the load wrench required for load-pose tracking; instead, the optimization changes the internal force components to meet the non-stopping requirement of the carriers, not altering the load dynamics. 
The proposed controller requires feedback of the load's state in the load wrench controller. In real applications, this may require an estimation algorithm, a sensorized payload, or sensors on board the carriers to track the load; on the other hand, no feedback of the cable forces is, instead, used by the method. The cable tension $T_i(t)$ in \eqref{eq:norm_epsilon} is computed as a function of the optimization variables as the norm of \eqref{eq:desired_cable_forces} and $\Pi_i$ is computed as its direction.

\section{Validation}

We validate the proposed framework through numerical simulations of a four-carrier system transporting a rigid load along a dynamic trajectory. The simulations account for full payload and cable dynamics, including non-idealities such as cable elasticity. Each cable is modeled as a massless spring-damper system, while each carrier is a double integrator controlled via a PID controller.  
System parameters and controller gains are summarized in Table~\ref{tab:sim_params_gains}.

\begin{figure}[t]
 \centering
 \includegraphics[width=\linewidth, trim={130 0 140 0}, clip]{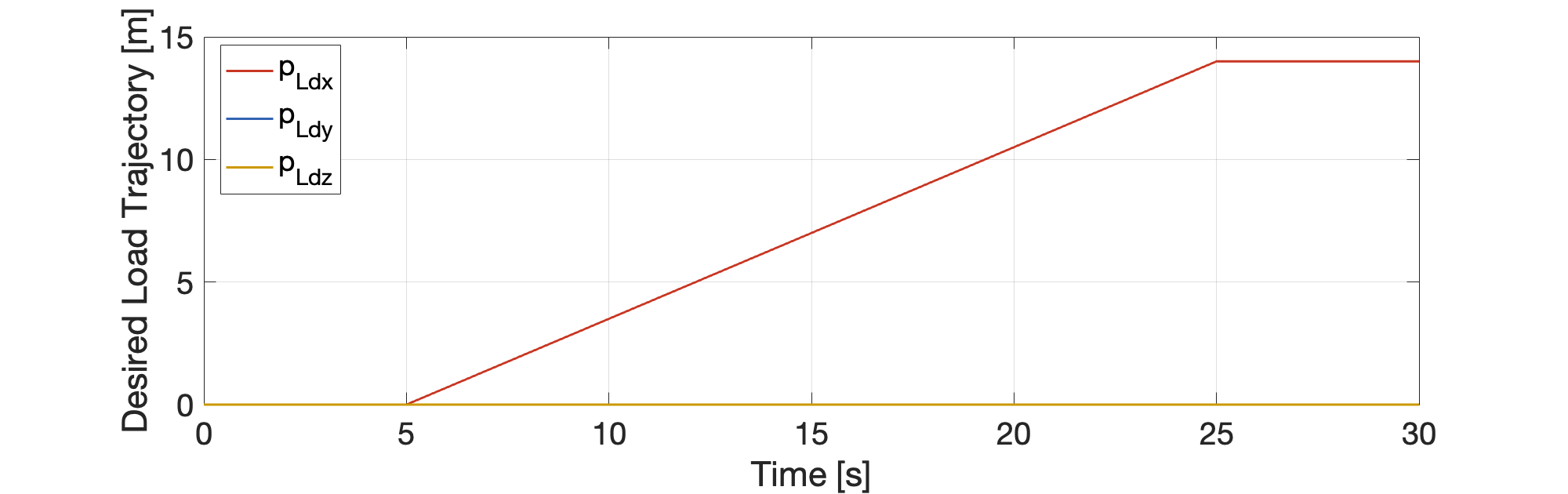}
 \caption{Components of the desired load trajectory. The load is initially static, then moves linearly along the x-axis, and finally remains static in its final position.}
 \label{fig:load_traj}
 \end{figure}

\begin{table}[t]
\centering
\caption{Simulation parameters and controller gains.}
\label{tab:sim_params_gains}
\begin{tabular}{l l c}
\toprule
\textbf{Symbol} & \textbf{Description} & \textbf{Value} \\
\midrule
$N$ & Number of carriers & $4$ \\
$K_d^c$ & Carrier derivative gain matrix & $1.5\, I_{3N}$ \\
$K_p^c$ & Carrier position gain matrix & $1000\, I_{3N}$ \\
$m_L$ & Load mass & $1.0~\mathrm{kg}$ \\
$J_L$ & Load inertia matrix & $0.01\, I_3~\mathrm{kg \cdot m^2}$ \\
$B_L$ & Load damping coefficient & $0.7~\mathrm{Ns/m}$ \\
$L_0$ & Cable rest length & $0.8~\mathrm{m}$ \\
$K_c$ & Cable stiffness & $500~\mathrm{N/m}$ \\
$B_c$ & Cable damping & $0.1~\mathrm{Ns/m}$ \\
$\varepsilon$ & Minimum carrier velocity & $0.2~\mathrm{m/s}$ \\
\midrule
$K_p$ & Position gain & $5$ \\
$K_v$ & Velocity gain & $2$ \\
$K_i$ & Integral gain & $0.9$ \\
$K_R$ & Orientation gain & $0.5$ \\
$K_\omega$ & Angular velocity gain & $0.06$ \\
$K_{iR}$ & Integral rotation gain & $0.1$ \\
\bottomrule
\end{tabular}
\end{table}

 \begin{figure}[t]
\centering
\subfloat[Without velocity optimization\label{fig:no_ott_traj}]{\includegraphics[width=0.9\linewidth]{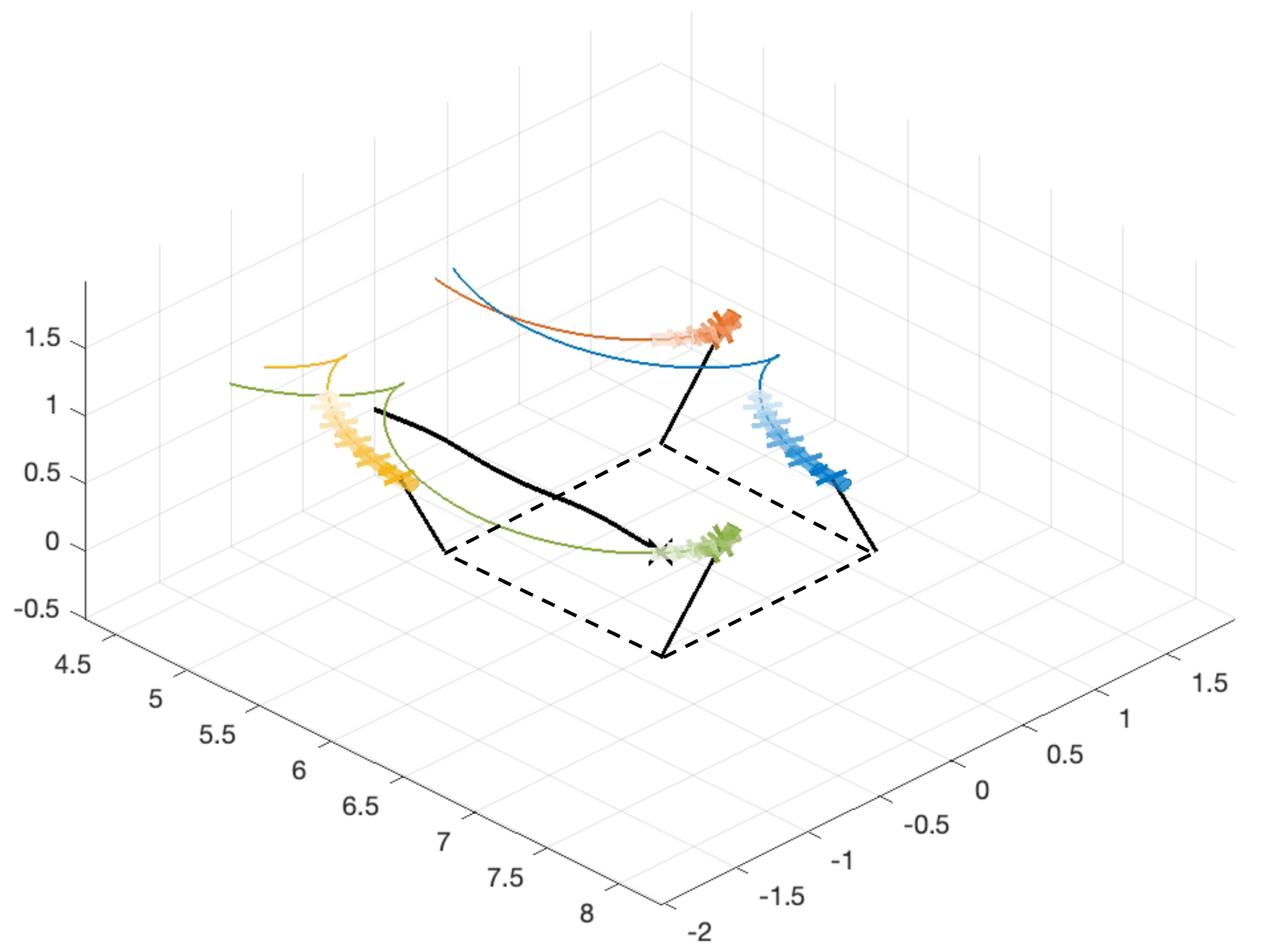}}\\
\subfloat[With velocity optimization \label{fig:si_ott_traj}]{\includegraphics[width=0.9\linewidth]{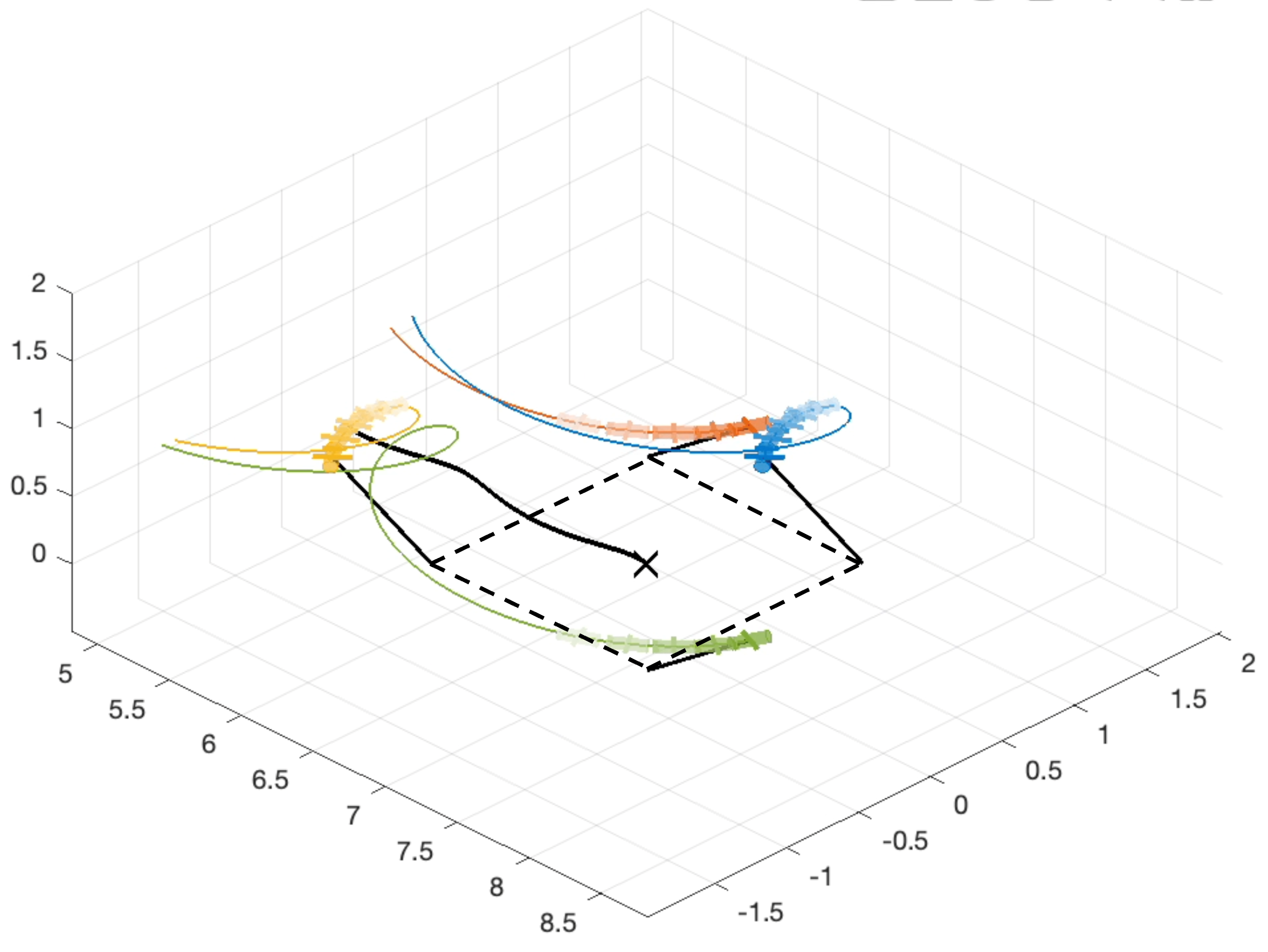}}
\caption{Zoom on the carrier trajectories (thin colored lines) without~(a) and with~(b) velocity optimization: in the first case, sharp turns result in the carriers stopping instantly. A schematic representation of the carrier is visible at the end of the trajectories as cross markers (higher transparency indicates a later time). The final cable configuration is represented by black lines, the load by a dashed square, and the load CoM by a cross. Axes are in meters. }
\end{figure}

\begin{figure}[!h]
\centering
\includegraphics[width=\linewidth, trim={40 250 90 300}, clip]{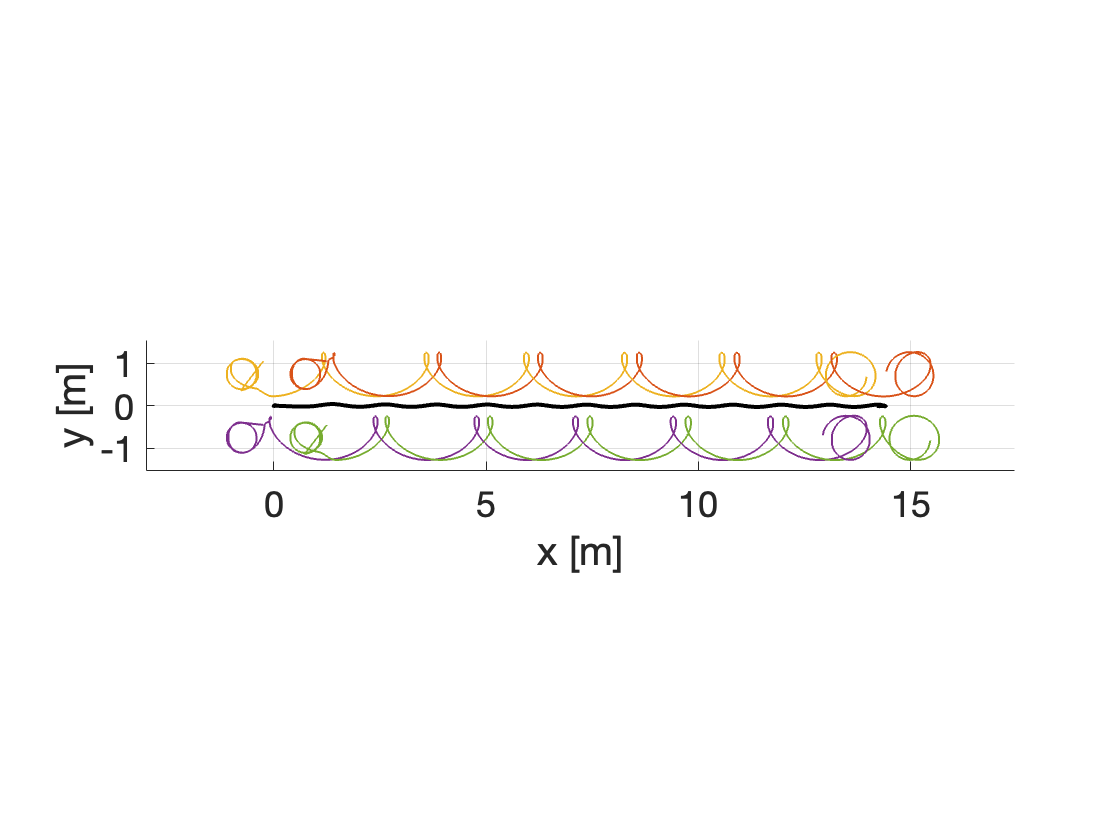}
\caption{Top-view of the trajectories, \emph{with} velocity optimization. The load position (black) starts from a static configuration, undergoes a motion phase, and returns to rest, while the carriers continue~moving to maintain feasibility, loitering above the load when it is static.}
\label{fig:si_ott_full_traj}
\end{figure}

\begin{figure}[t]
\centering
\subfloat[Load position error]{\includegraphics[width=0.7\linewidth,trim=10 0 30 10, clip]{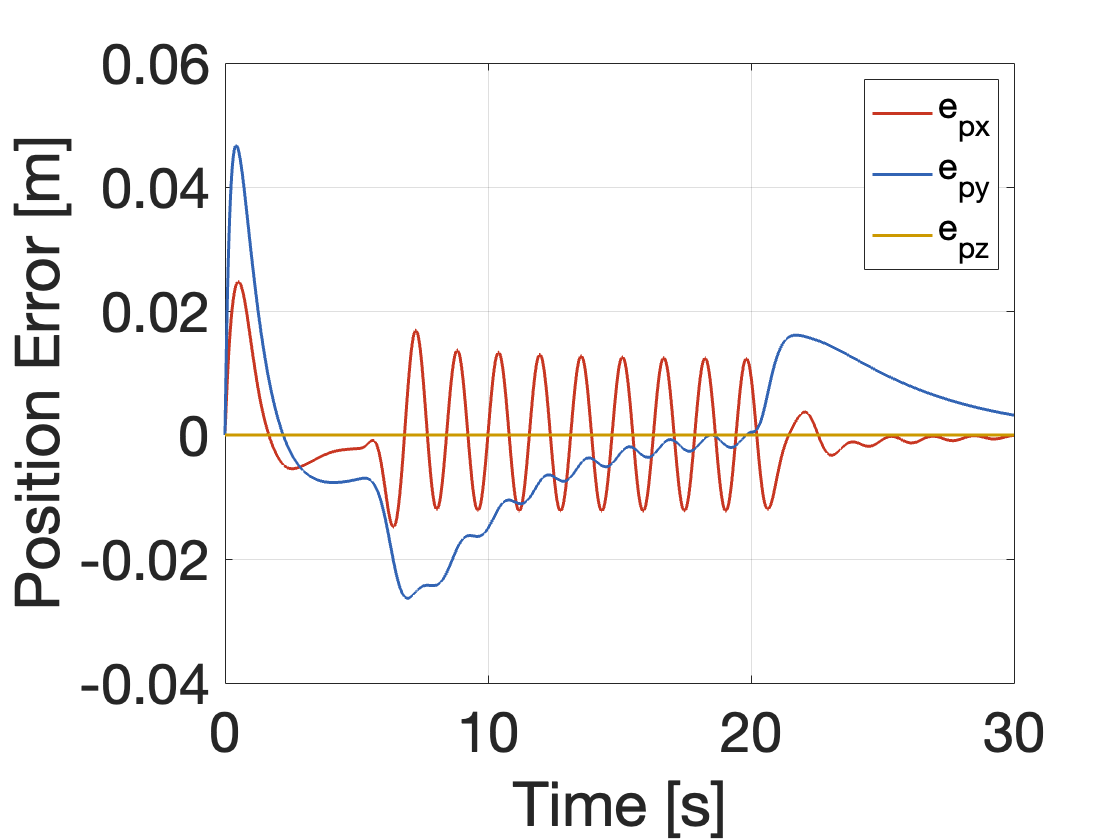}}
\hfill
\subfloat[Load orientation error]{\includegraphics[width=0.7\linewidth,trim=10 0 30 10, clip]{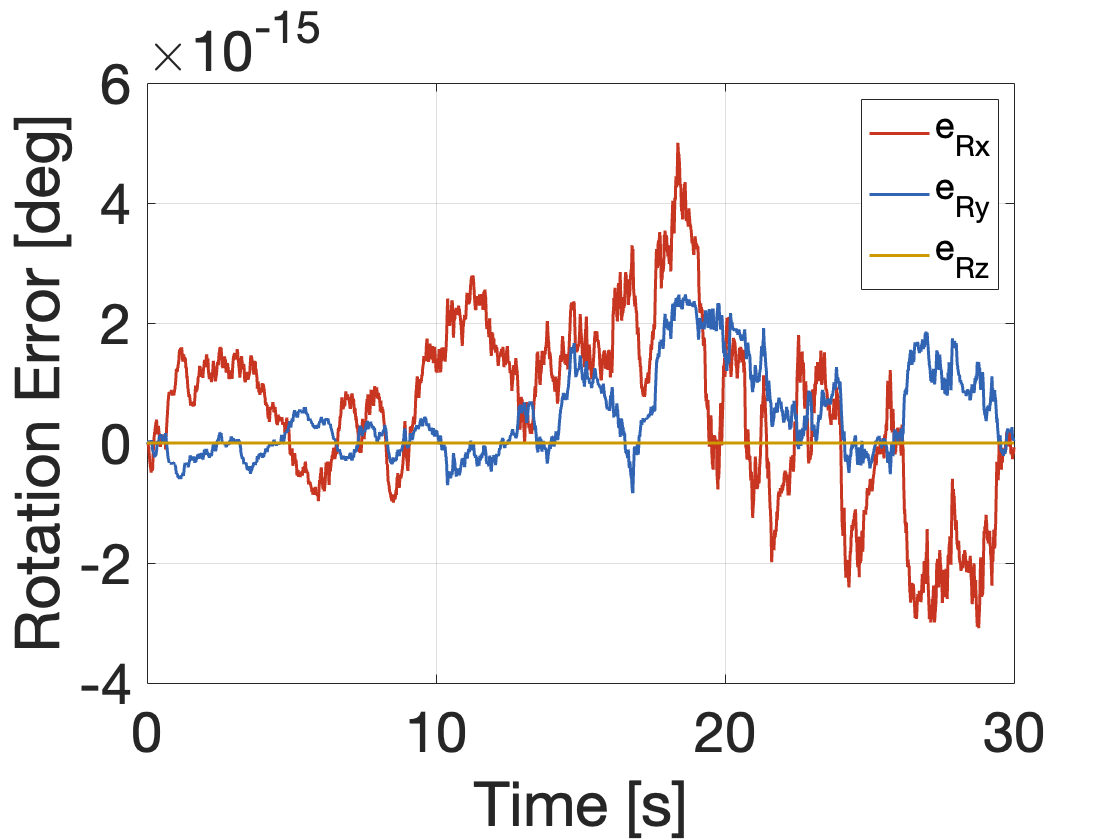}}
\caption{Load tracking errors without velocity optimization.}
\label{fig:no_ott_errors}
\end{figure}

\begin{figure}[!h]
\centering
\includegraphics[width=0.75\linewidth,trim=10 0 30 10, clip]{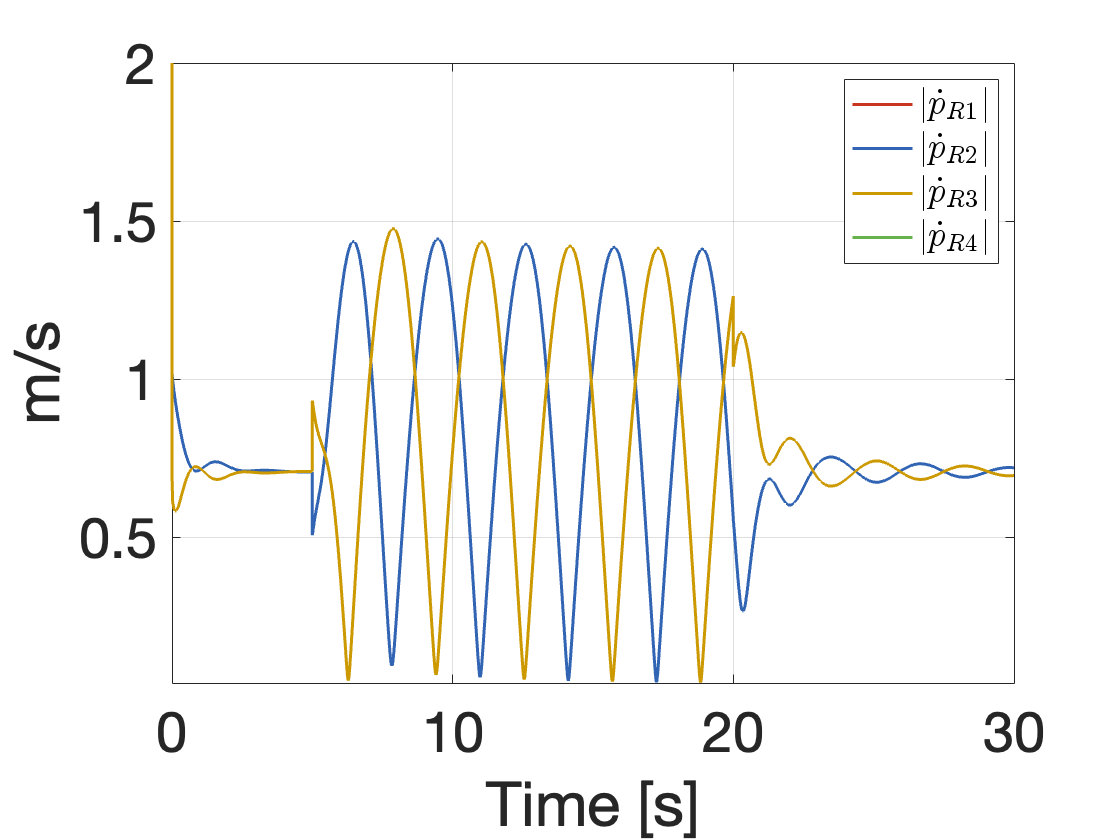}
\caption{Carrier velocity norms without optimization. Carrier velocities drop to zero during sharp turns, violating fixed-wing constraints.}
\label{fig:no_ott_v}
\end{figure}

The load desired trajectory, represented in Figure~\ref{fig:load_traj},  consists of a static initial position, a linear displacement to a target location, and a final static phase. This trajectory highlights situations where carrier velocity may drop to zero if non-optimized internal-force trajectories are used.


Figures~\ref{fig:no_ott_traj} and \ref{fig:no_ott_errors}-\ref{fig:no_ott_v} show the system behavior under the proposed controller \textit{without} internal force optimization. During the translational motion of the load, the carriers perform sharp inverse turns to maintain the load trajectory, leading to temporary drops in their velocity norms to zero.
While load tracking remains reasonably accurate, the near-zero carrier velocities demonstrate the practical limitations of non-optimized trajectories for non-stopping UAVs.
\begin{figure}[!h]
\centering
\subfloat[Load position error]{\includegraphics[width=0.7\linewidth,trim=10 0 30 10, clip]{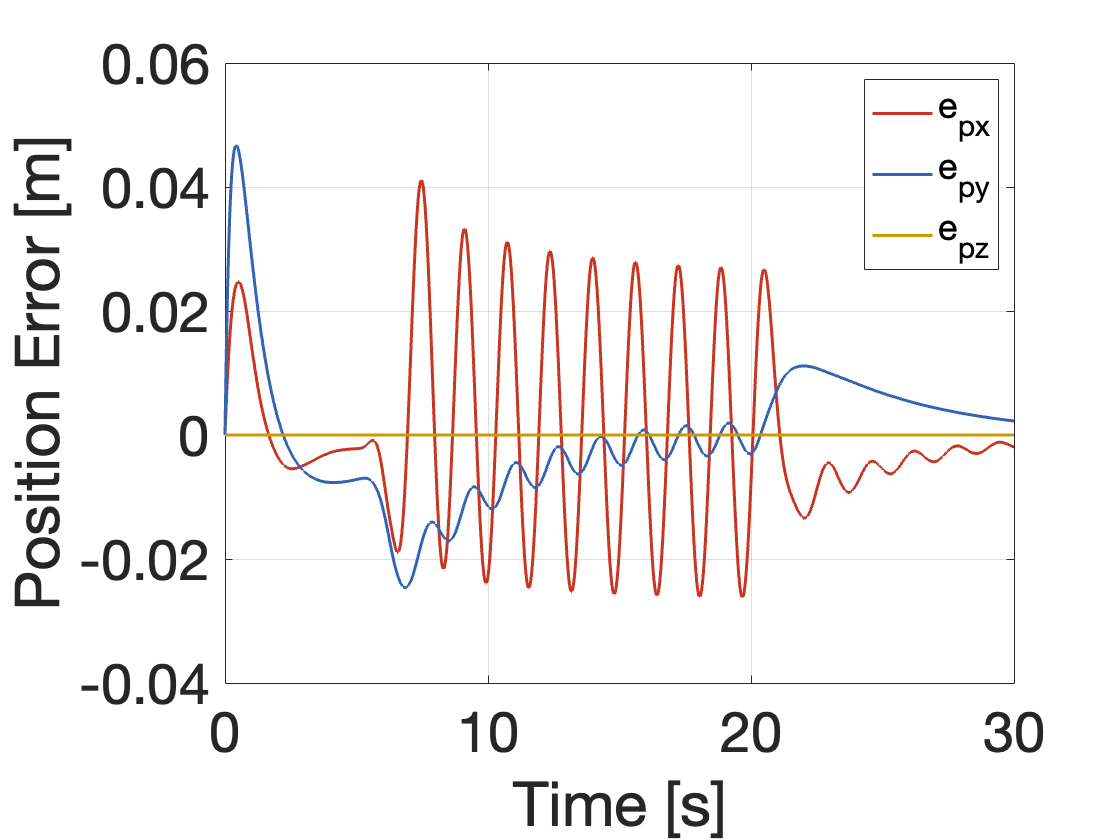}}
\hfill
\subfloat[Load orientation error]{\includegraphics[width=0.7\linewidth,trim=10 0 30 10, clip]{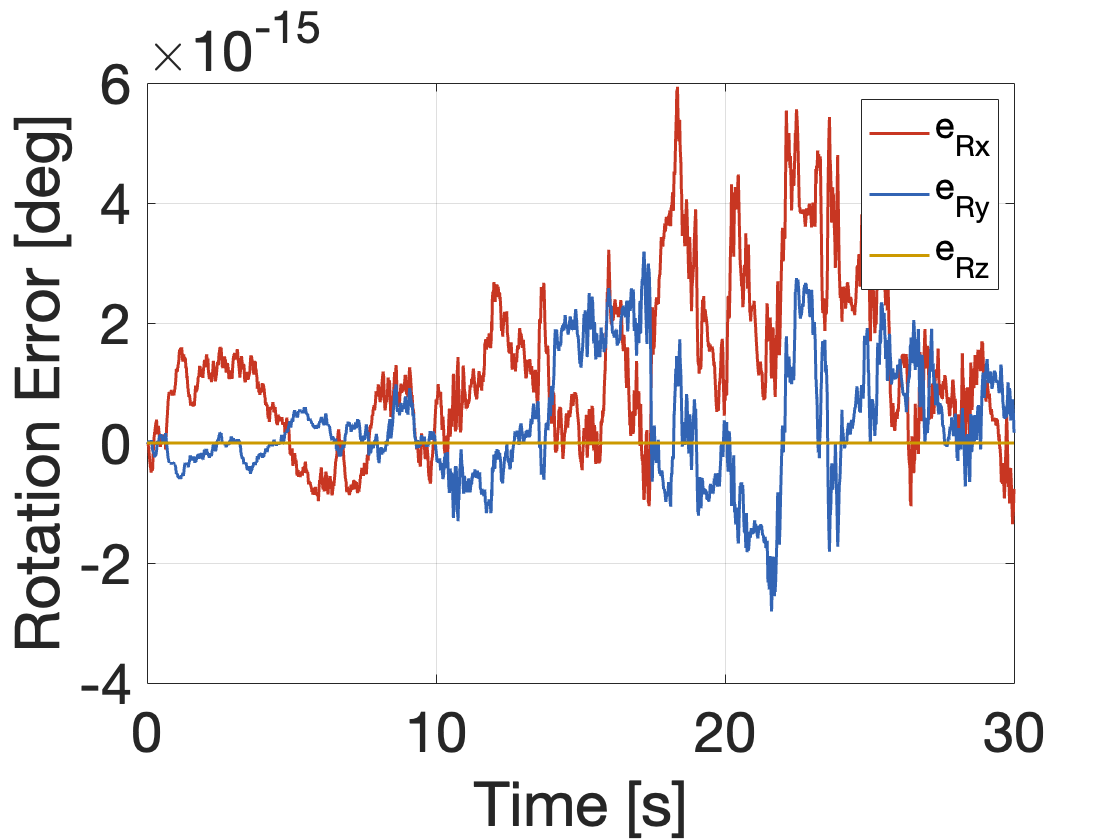}}
\caption{Load tracking errors with velocity optimization.}
\label{fig:si_ott_errors}
\end{figure}
\begin{figure}[!h]
\centering
\includegraphics[width=0.75\linewidth,trim=10 0 30 10, clip]{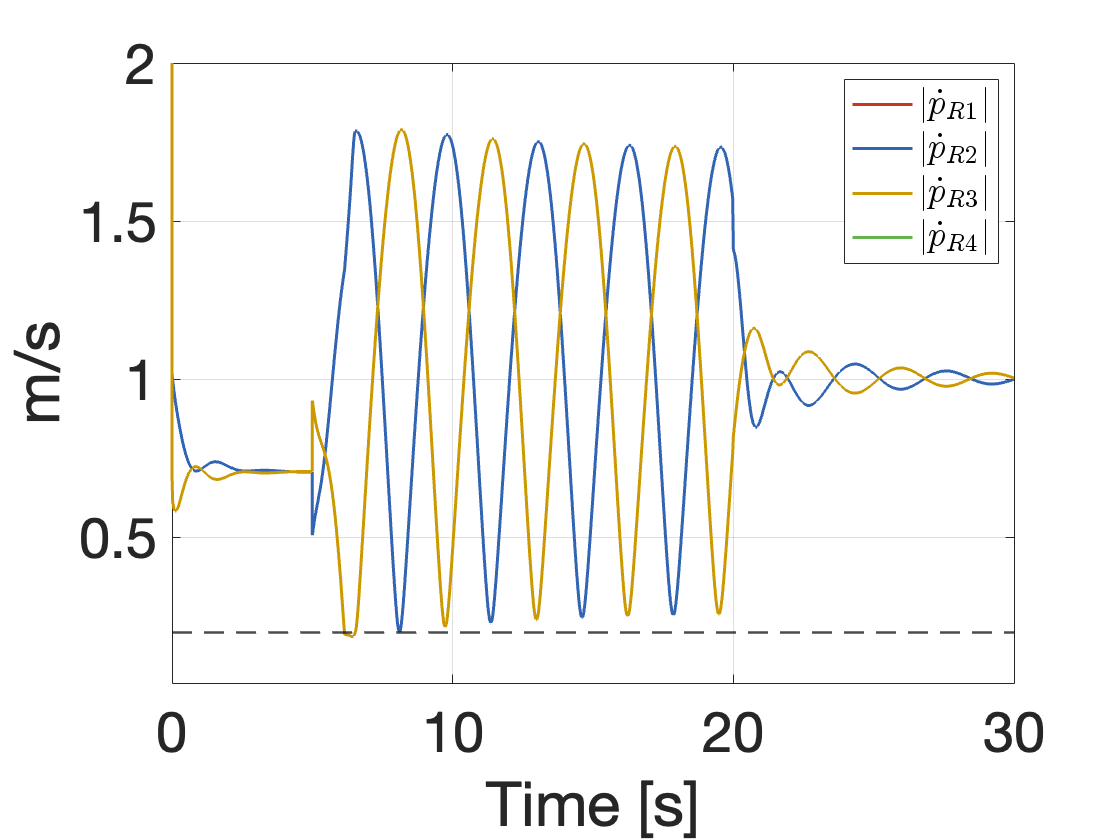}
\caption{Carrier velocity norms with optimization. Carrier velocities maintain a norm above the imposed threshold  $\varepsilon = 0.2~\mathrm{m/s}$.}
\label{fig:si_ott_v}
\end{figure}

\begin{figure*}[t!]
    \centering
    \includegraphics[width=0.5\linewidth]{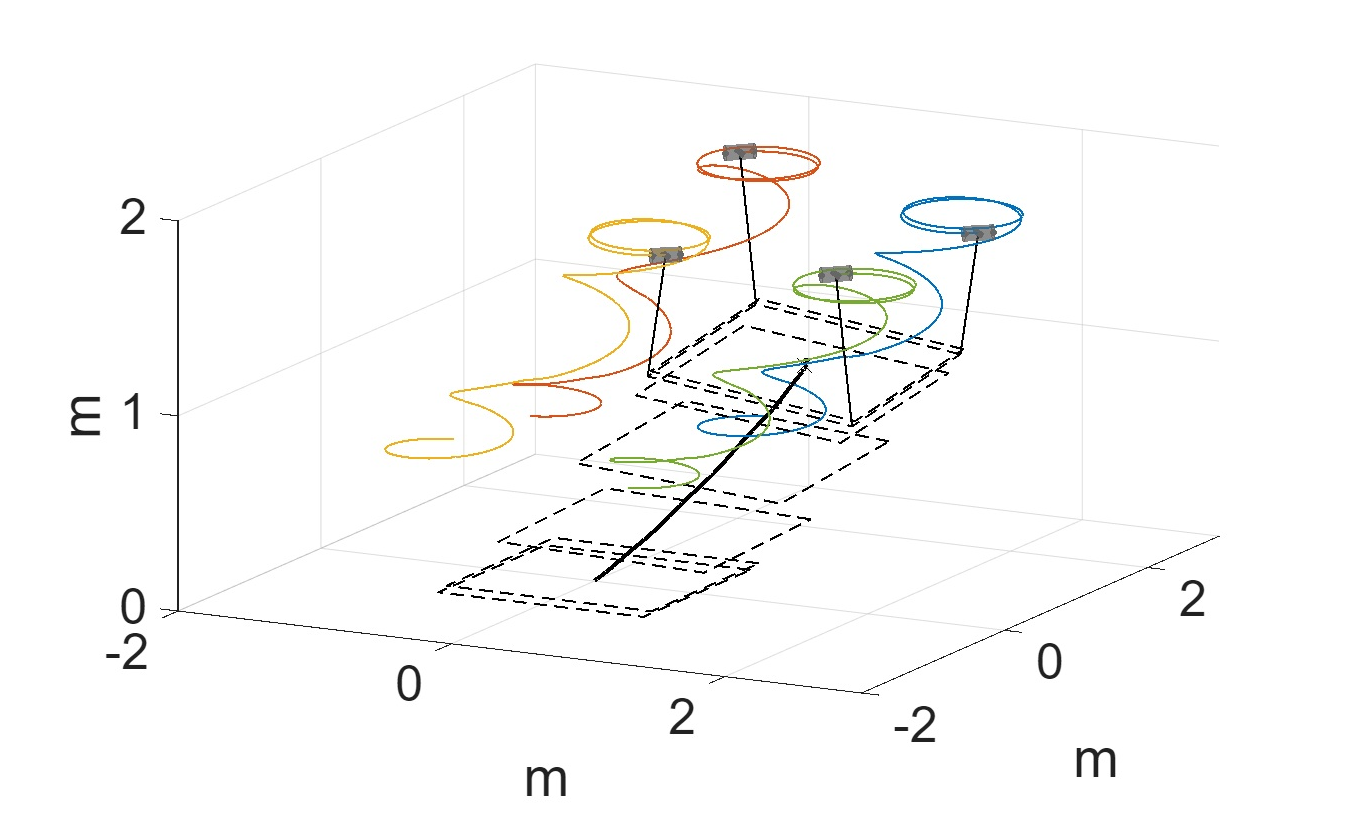}\includegraphics[width=0.5\linewidth]{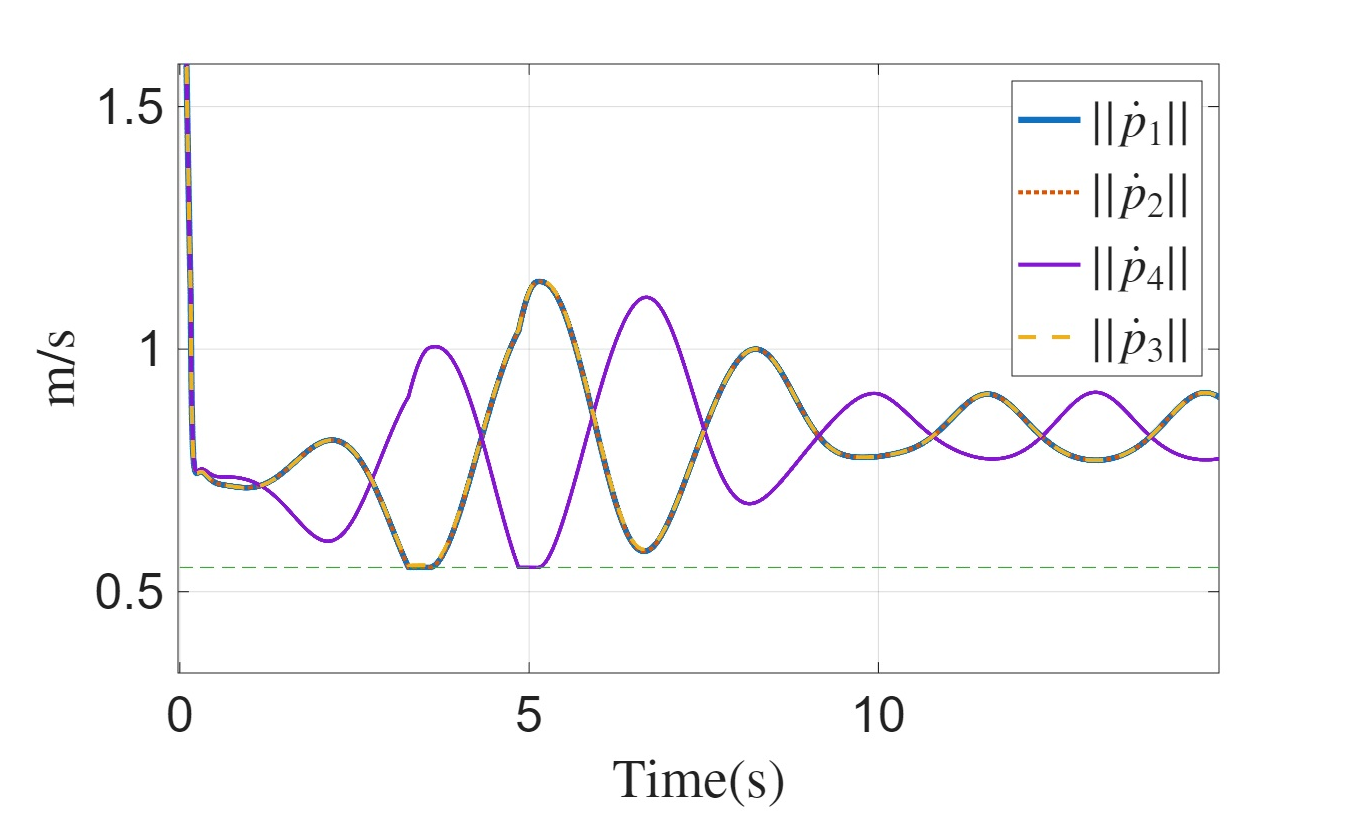}\\
    \includegraphics[width=0.5\linewidth]{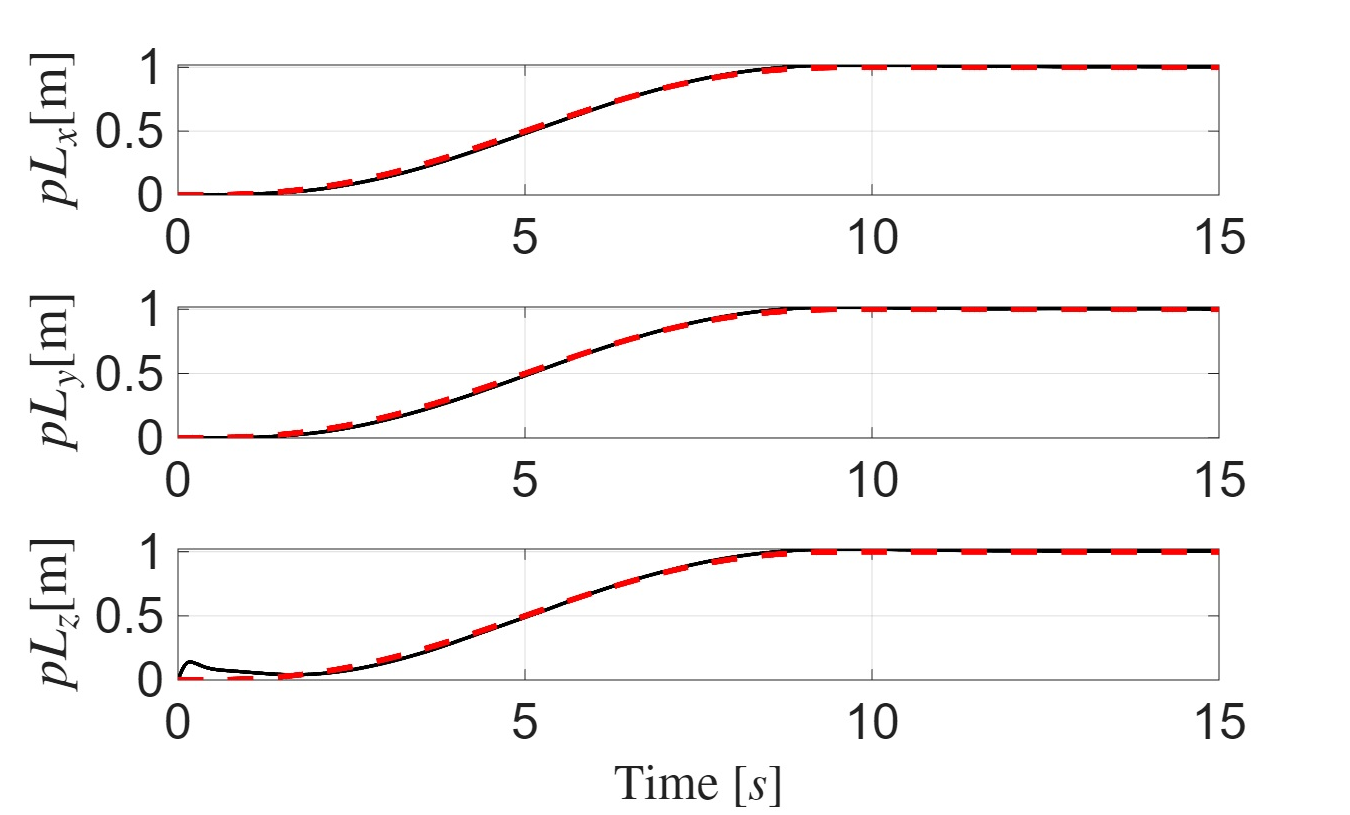}\includegraphics[width=0.5\linewidth]{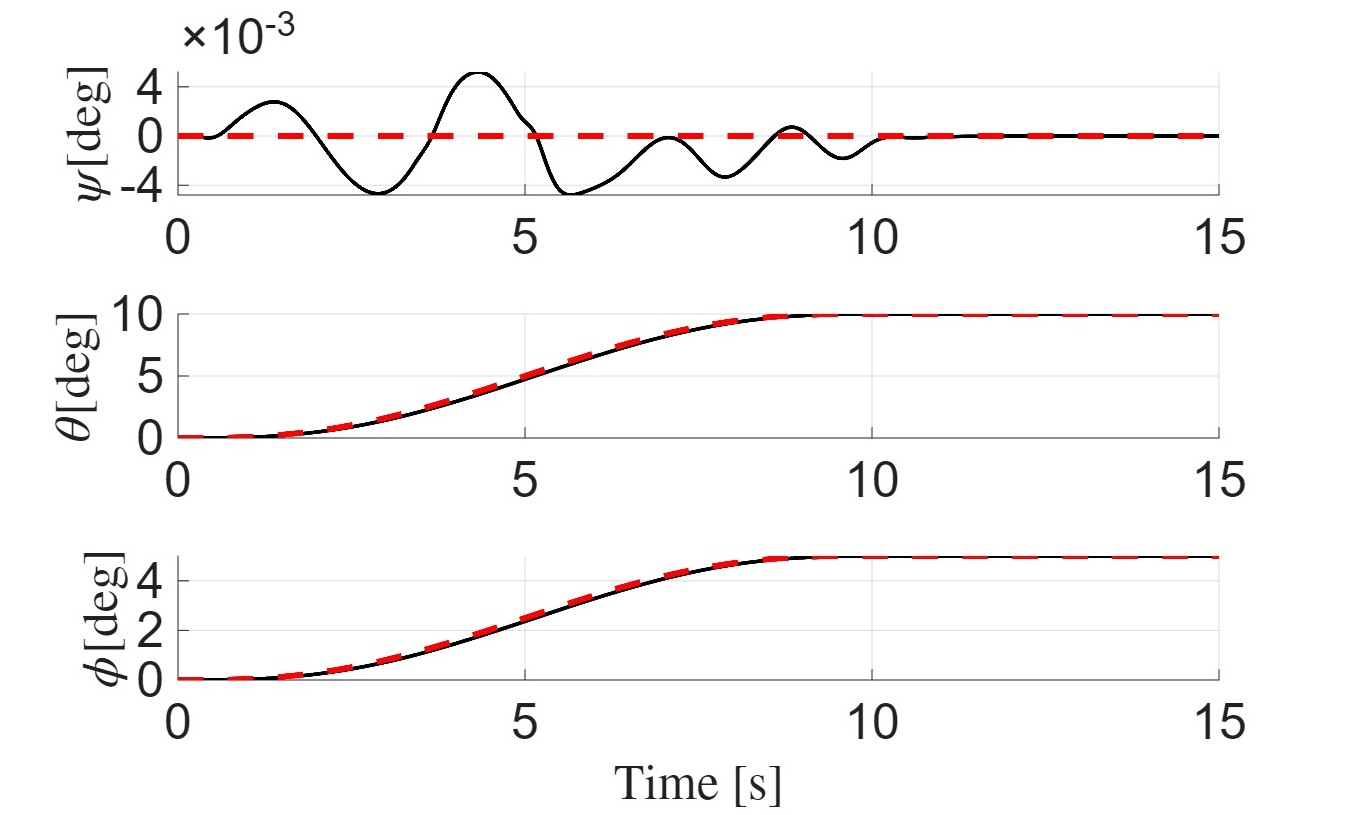}
    \caption{Simulation results for the desired load trajectory shown at the bottom. Both the position (bottom-left) and orientation (bottom-right) of the load follow a component-wise rest-to-rest quintic polynomial trajectory. 
    At the top-left, a schematic representation of the system, where different snapshots of the load are depicted in dashed black lines; the solid black line is the load CoM position over time, and the colored lines are the positions of the carriers.
    At the top-right, the norm of the carriers' velocities is modified not to exceed the imposed lower-bound, in a dashed green line. 
    }
    \label{fig:complete_traj}
\end{figure*}


Figures~\ref{fig:si_ott_traj} and \ref{fig:si_ott_full_traj}-\ref{fig:si_ott_errors} show the carriers' and load's motion \textit{with} the proposed velocity optimization layer that adjusts the amplitude and frequency of internal force trajectories to enforce $\|\dot p_{Ri}\| \ge \varepsilon$ for all carriers. Note how the norm of the carrier velocities hits the lower bound without exceeding it, thanks to the proposed control law.

In Figure~\ref{fig:complete_traj}, another simulation result is shown, with ${\varepsilon=0.55}$ and quintic polynomial trajectories in the load's position and attitude components.

Each coordinate of the load pose is required to move by 1 m in 10 s, the pitch angle by 10 degrees, and the roll angle by 5 degrees. Then, the pose of the load is asked to remain constant for 5 seconds. As can be seen in Figure~\ref{fig:complete_traj}, the load pose follows the reference value. After around 3s, the norms of the velocity of carriers 1, 2, and 3 hit the lower bound, and the optimization algorithm successfully prevents them from exceeding it; the same happens to the norm of the velocity of carrier 4 at a time equal to about 5 seconds. In the meantime, the load trajectory tracking is not perturbed.

The simulation results show that the proposed method allows the carriers to keep their forward velocity while also manipulating the suspended load along a trajectory composed of static and dynamic parts. From the comparison, the following aspects emerge.
\begin{itemize}
    \item \textit{Carrier velocity enforcement:} the optimization layer ensures strictly non-zero velocity norms of the carriers, respecting non-stopping constraints and keeping the velocity norms above the minimum threshold of $\varepsilon = 0.2$~m/s throughout the simulation.
    \item \textit{Load tracking performance:}  Position and orientation errors remain bounded and converge to zero when the load desired trajectory is static, demonstrating that enforcing minimum velocity does not compromise load stability. We observe slightly higher position error of the load in the optimized case in Figure~\ref{fig:si_ott_errors} than in the non-optimized one in Figure~\ref{fig:no_ott_errors}, with the mean value of $||e_p||$ equal to 0.015 m in the former case and 0.013 m in the latter. This may be due to the higher velocities reached by the carriers, which excite the visco-elastic simulated cables. 
\end{itemize}


\section{Conclusions and Future Work}

We introduced a closed-loop control framework for cooperative transport of a cable-suspended load using non-stopping carriers, constrained to maintain non-zero velocities. The method combines a feedback wrench controller with an optimization-based internal-force generation layer to enable non-stop carrier trajectories while achieving load tracking. Simulation results demonstrated successful load trajectory tracking while avoiding carriers' stagnation. 

Future work will overcome the simplified carrier dynamics, refining the trajectory of the carriers by accounting for vehicle-specific dynamics and input saturation. Potential cases where the optimization may fail to find a solution at certain instants because of incompatible constraints will be handled. The proposed online optimization reactively changes the internal forces; in the future, a receding-horizon approach could be explored. The feasibility of the online implementation will be further studied. Moreover, we will work on a thorough experimental validation on multirotor UAVs emulating fixed-wing constraints. Ultimately, we will consider the implementation on actual convertible UAVs, handling the necessary mechatronics adaptations needed to host the cable and load, as well as the load pick-up and release. Online load-state estimation will be integrated.  
\newpage
\printbibliography[title={References}]

@inproceedings{saengphet2016conceptual,
  title={Conceptual design of fixed wing-VTOL UAV for AED transport},
  author={Saengphet, Watcharapol and Thumthae, Chalothorn},
  booktitle={The 7th TSME International Conference on Mechanical Engineering},
  pages={1--16},
  year={2016}
}

@article{zhang2024design,
  title={Design and control of an ultra-low-cost logistic delivery fixed-wing UAV},
  author={Zhang, Yixuan and Zhao, Qinyang and Mao, Peifu and Bai, Qiaofeng and Li, Fuzhong and Pavlova, Svitlana},
  journal={Applied Sciences},
  volume={14},
  number={11},
  pages={4358},
  year={2024},
  publisher={MDPI}
}

@ARTICLE{8825990,
  author={Xian, Bin and Wang, Shizhang and Yang, Sen},
  journal={IEEE Transactions on Industrial Electronics}, 
  title={An Online Trajectory Planning Approach for a Quadrotor UAV With a Slung Payload}, 
  year={2020},
  volume={67},
  number={8},
  pages={6669-6678},
  doi={10.1109/TIE.2019.2938493}}

@article{leutenegger2016flying,
  title={Flying robots},
  author={Leutenegger, Stefan and H{\"u}rzeler, Christoph and Stowers, Amanda K and Alexis, Kostas and Achtelik, Markus W and Lentink, David and Oh, Paul Y and Siegwart, Roland},
  journal={Springer Handbook of Robotics},
  pages={623--670},
  year={2016},
  publisher={Springer}
}

@article{estevez2024review,
  title={Review of aerial transportation of suspended-cable payloads with quadrotors},
  author={Estevez, Julian and Garate, Gorka and Lopez-Guede, Jose Manuel and Larrea, Mikel},
  journal={Drones},
  volume={8},
  number={2},
  pages={35},
  year={2024},
  publisher={MDPI}
}

@article{li2021cooperative,
  title={Cooperative transportation of cable suspended payloads with mavs using monocular vision and inertial sensing},
  author={Li, Guanrui and Ge, Rundong and Loianno, Giuseppe},
  journal={IEEE Robotics and Automation Letters},
  volume={6},
  number={3},
  pages={5316--5323},
  year={2021},
  publisher={IEEE}
}

@article{gabellieri2023equilibria,
  title={Equilibria, stability, and sensitivity for the aerial suspended beam robotic system subject to parameter uncertainty},
  author={Gabellieri, Chiara and Tognon, Marco and Sanalitro, Dario and Franchi, Antonio},
  journal={IEEE Transactions on Robotics},
  volume={39},
  number={5},
  pages={3977--3993},
  year={2023},
  publisher={IEEE}
}

@inproceedings{gabellieri2024existence,
  title={On the Existence of Static Equilibria of a Cable-Suspended Load with Non-stopping Flying Carriers},
  author={Gabellieri, Chiara and Franchi, Antonio},
  booktitle={2024 International Conference on Unmanned Aircraft Systems (ICUAS)},
  pages={638--644},
  year={2024},
  organization={IEEE}
}

@inproceedings{Sreenath2013,
  title={Trajectory generation and control of a quadrotor with a cable-suspended load---a differentially-flat hybrid system},
  author={Sreenath, K. and Michael, N. and Kumar, V.},
  booktitle={2013 IEEE International Conference on Robotics and Automation (ICRA)},
  pages={4888--4895},
  year={2013},
  organization={IEEE}
}

@inproceedings{Pereira2016,
  title={Slung load transportation with a single aerial vehicle and disturbance removal},
  author={Pereira, P.O. and Herzog, M. and Dimarogonas, D.V.},
  booktitle={2016 24th Mediterranean Conference on Control and Automation (MED)},
  pages={671--676},
  year={2016},
  organization={IEEE}
}

@inproceedings{Chen2019,
  title={Cooperative transportation of cable-suspended slender payload using two quadrotors},
  author={Chen, T. and Shan, J.},
  booktitle={2019 IEEE International Conference on Unmanned Systems (ICUS)},
  pages={432--437},
  year={2019},
  organization={IEEE}
}

@inproceedings{Masone2016,
  title={Cooperative transportation of a payload using quadrotors: A reconfigurable cable-driven parallel robot},
  author={Masone, C. and B{\"u}lthoff, H.H. and Stegagno, P.},
  booktitle={2016 IEEE/RSJ International Conference on Intelligent Robots and Systems (IROS)},
  pages={1623--1630},
  year={2016},
  doi={10.1109/IROS.2016.7759262},
  organization={IEEE}
}

@article{michael2011cooperative,
  title={Cooperative manipulation and transportation with aerial robots},
  author={Michael, Nathan and Fink, Jonathan and Kumar, Vijay},
  journal={Autonomous Robots},
  volume={30},
  pages={73--86},
  year={2011},
  publisher={Springer}
}

@article{gabellieri2025coordinated,
  title={Coordinated Trajectories for Non-stop Flying Carriers Holding a Cable-Suspended Load},
  author={Gabellieri, Chiara and Franchi, Antonio},
  journal={arXiv preprint arXiv:2503.03481},
  year={2025}
}

\end{document}